\documentclass[sigconf]{acmart}
\AtBeginDocument{%
  }

\setcopyright{acmlicensed}
\copyrightyear{2026}
\acmYear{2026}
\setcopyright{cc}
\setcctype{by}
\acmConference[CIKM '26]{Proceedings of the 35th ACM International Conference on Information and Knowledge Management}{November 07--11, 2026}{Rome, Italy}
\acmBooktitle{Proceedings of the 35th ACM International Conference on Information and Knowledge Management (CIKM '26), November 07--11, 2026, Rome, Italy}
\acmDOI{10.1145/3799682.3840077}
\acmISBN{979-8-4007-2539-5/2026/11}

\begin{document}

\title{Verification-Guided Specification Synthesis with Large Language Models for Intrusion Detection Rules}

\author{Kohei Yamamoto}
\email{yamamoto26@mm.doshisha.ac.jp}
\affiliation{%
  \institution{Doshisha University}
  \streetaddress{1--3 Tatara Miyakodani}
  \city{Kyotanabe-shi}
  \state{Kyoto}
  \country{Japan}
}

\author{Marie Katsurai}
\email{katsurai@mm.doshisha.ac.jp}
\affiliation{%
  \institution{Doshisha University}
  \streetaddress{1--3 Tatara Miyakodani}
  \city{Kyotanabe-shi}
  \state{Kyoto}
  \country{Japan}
}

\renewcommand{\shortauthors}{Yamamoto and Katsurai}

\begin{abstract}
Attacks against Internet-connected IoT devices continue to increase; however, transforming observed attack traffic into deployable intrusion detection system (IDS) rules remains largely a manual process. Recent studies have explored using large language models (LLMs) to generate IDS rules; nonetheless, existing approaches often require auxiliary information beyond observed traffic or generate rules without validating their detection logic against benign traffic.
This study presents a verification-guided specification synthesis framework for generating Suricata rules directly from HTTP request traces. Instead of having an LLM generate IDS rules in a single step, an LLM first identifies a vulnerable parameter and synthesizes a semantic detection specification. These specifications are iteratively refined through counterexample-guided inductive synthesis (CEGIS), in which benign traffic samples serve as counterexamples during synthesis and verification. Verified specifications are then deterministically compiled into Suricata rules.  
Experiments on 281 real-world CVEs and benign traffic collected from real IoT devices show that the proposed method achieves a detection rate of 81.5\% while maintaining a false positive rate of 0.0\%. An ablation study also demonstrates that CEGIS-based verification improves detection performance while maintaining a low false positive rate.

\end{abstract}





\maketitle

\section{Introduction}
\label{sec:intro}

Large-scale attacks on Internet-connected IoT devices continue to be observed~\cite{ciciot23}. Since many vulnerable IoT devices have reached end-of-life and no longer receive firmware updates, network-based protection through intrusion detection systems (IDSs) remains an important defense mechanism. In practice, transforming newly observed attack traffic into deployable IDS rules remains largely a manual process requiring substantial security expertise and rule-authoring effort~\cite{scarfone2007guide,hoque2025datadriven}, creating a delay between attack observation and rule deployment.

Recent advances in large language models (LLMs) have demonstrated their potential to assist security analysts in extracting knowledge from unstructured security observations~\cite{deng2024pentestgpt}. Several studies have explored using LLMs with attack-related sources, such as vulnerability descriptions, cyber threat intelligence (CTI) reports, exploit code, and attack traces~\cite{falcon,gridai,rulexploit} to construct deployable detection rules for signature-based IDSs, such as Suricata\footnote{https://suricata.io/}.
However, although attack requests themselves are easily available from honeypots and monitoring systems~\cite{iotpot16,torchlight},
the auxiliary information required by existing approaches is often unavailable immediately after an attack is observed.
Furthermore, because LLMs are typically prompted to generate Suricata rules in a single step, without validating the generated detection logic against benign traffic~\cite{gridai}, generated rules may overfit the observed attack examples and produce false positives on legitimate communications.
In operational settings, IDS rules are expected not only to detect attacks but also to avoid triggering on benign traffic. Therefore, before a rule is deployed, the underlying detection condition should be verified against examples that represent legitimate behavior.

This paper presents a verification-guided framework that generates intrusion detection rules directly from HTTP request traces. Instead of prompting an LLM to generate Suricata syntax, we use the LLM to synthesize semantic detection specifications that characterize attack behavior, including vulnerable parameters and attack mechanisms. The specifications are iteratively refined through counterexample-guided inductive synthesis (CEGIS)~\cite{cegis}, where benign traffic samples serve as counterexamples during synthesis and verification.
Only verified specifications are compiled into deployable Suricata rules.
Experiments on 281 real-world CVEs and benign traffic collected from real IoT environments demonstrate that the proposed framework achieves high attack detection performance while effectively suppressing false positives. Furthermore, comparisons with direct LLM-based rule generation methods and ablation studies confirm the effectiveness of separating specification synthesis from rule compilation and of incorporating counterexample-guided verification.

\begin{figure*}[t]
  \centering
  \includegraphics[width=0.94\linewidth]{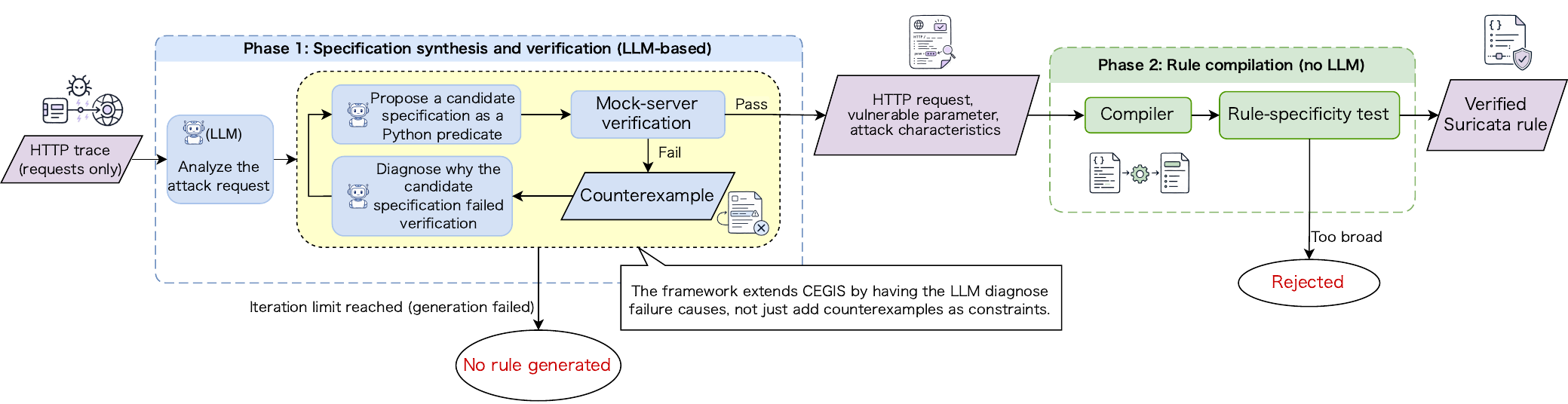}
  \vspace{-0.2cm}
  \caption{Overview of the proposed method.}
  \label{fig:overview}
\end{figure*}

The main contributions of this paper are twofold:
(i) we propose a novel LLM-based framework that extracts vulnerable parameters and attack mechanisms directly from HTTP request traces and refines synthesized specifications through counterexample-guided verification before compiling them into Suricata rules;
and
(ii) we introduce an evaluation protocol that separates benign traffic used for specification synthesis from benign traffic used for evaluation and evaluates all methods using identical attack traces, IDS settings, and LLM configurations. The experiments reveal distinct limitations across approaches: conventional non-LLM methods frequently fail to generate rules that correctly detect their target attacks, while LLM-based methods often produce rules that fail to trigger correctly in Suricata.

\section{Related Work}
\label{sec:related}
Automatic IDS rule generation has been studied using a variety of signature synthesis techniques. 
For example, Syrius~\cite{syrius} prepares overly specific seed rules and refines them using benign traffic, whereas CMIRGen~\cite{cmirgen} and AutoCombo~\cite{autocombo} generate signatures through clustering, rule mining, and frequent-pattern discovery techniques. 
Because these approaches derive signatures from statistical differences between attack and benign traffic, they may fail to identify attacks whose syntax closely resembles legitimate communications. 


Among recent LLM-based approaches, FALCON~\cite{falcon} generates IDS and malware detection rules from CTI reports, while RuleXploit~\cite{rulexploit} generates Suricata rules from exploit code and iteratively repairs them using attack traffic.
GRIDAI~\cite{gridai} directly generates and repairs Suricata rules from HTTP attack requests using multiple cooperating LLM agents; this is the most closely related to ours in terms of inputs.
RuleForge~\cite{ruleforge} generates detection rules from templates in an open-source vulnerability scanner, while UniRule~\cite{unirule} generates detection rules from natural-language requirements.
These LLM-based methods generate detection rules in a single step, without incorporating benign traffic as counterexamples during rule construction.

Our study differs from the above studies in several aspects. The proposed method requires neither CTI reports, exploit code, attacker labels, nor labeled benign corpora. It operates directly on HTTP request traces. Instead of generating Suricata rules in a single step, an LLM synthesizes semantic detection specifications, which are refined and verified using benign traffic as counterexamples through CEGIS before being compiled into Suricata rules.

\section{Proposed Method}
\label{sec:proposed}
In operational IDS settings, attack requests are often the only artifacts immediately available after an attack is observed. Thus, the proposed framework operates directly on HTTP request traces and generates verified Suricata rules without requiring HTTP responses, firmware images, or vulnerability descriptions. As illustrated in Figure~\ref{fig:overview}, the framework consists of two phases: (1) specification synthesis and verification and (2) rule compilation.
The key design principle is to separate specification synthesis from rule compilation. LLMs are used to derive detection specifications from HTTP attack requests, while a compiler deterministically translates the HTTP request and the corresponding analysis result into deployable Suricata rules. This separation avoids relying on LLMs for generating syntax-sensitive Suricata rules and enables verification at the specification level before rule compilation.

\subsection{Verification-Guided Specification Synthesis}
\label{ssec:stage1}
The LLM first analyzes an HTTP attack request and identifies a parameter that is likely to serve as the attack entry point. Many requests contain both security-relevant and irrelevant fields, and directly generating rules from the entire request often leads to overly general detection logic. Therefore, the parameter is selected according to its likely operational role within the target service, such as command execution, data access, path construction, or access control. The analysis relies solely on the observed HTTP request and does not require external vulnerability information. For requests that do not contain a plausible attack entry point, no rule is generated.

After identifying a vulnerable parameter, the LLM proposes a candidate semantic detection specification. Specifically, rather than generating IDS rules in a single step, the LLM is instructed to characterize the conditions under which a parameter value should be considered malicious. To guide synthesis, benign values collected from public IoT traffic datasets (e.g.,~\cite{wannigama2025unsw,ciciot23}) and the attack value observed in the target request are provided as negative and positive examples, respectively. Specifically, we choose representative values of a parameter by applying $k$-medoids clustering to its values in benign traffic data. In our preliminary experiments, the choice of $k$ does not significantly change the overall performance. The LLM then generates a Python predicate that distinguishes the attack value from benign ones. This intermediate representation captures the attack semantics while remaining independent of any IDS-specific syntax.

The generated specification is verified using the CEGIS strategy, a program-synthesis framework in which candidate solutions are iteratively refined using counterexamples produced by a verifier. In the proposed framework, the LLM acts as the synthesizer by generating candidate detection specifications, while a verifier evaluates them on a mock server that emulates the target service. The mock server is a lightweight web server generated by the LLM; it receives attack requests at the same endpoint paths as the target service and processes the identified vulnerable parameter, eliminating the need to access real vulnerable devices. This setup enables the framework to determine whether the identified parameter influences server behavior and whether the synthesized specification distinguishes attack inputs from benign values.
Verification consists of four checks: (1) the attack request elicits a valid server response, (2) modifying the identified parameter changes server behavior, (3) multiple benign values extracted from real traffic do not satisfy the specification, and (4) the processing of the vulnerable parameter is observable in server logs. The first two checks confirm that the attack affects the emulated service as intended, the third ensures that benign values are not falsely classified as malicious, and the fourth verifies that the parameter is actually processed by the server. Any failed check is treated as a counterexample and returned to the synthesizer.

When a specification fails verification, the framework diagnoses the cause of the failure and then generates a revised specification accordingly. Candidate specifications, counterexamples, and diagnostic results are stored in memory and reused in subsequent iterations to avoid repeating previous mistakes. 
The synthesis-verification loop is repeated for at most 13 iterations, a limit determined empirically in this study. If no valid specification is obtained within the iteration limit, the attack is discarded and no Suricata rule is generated.
Our method favors explicit non-generation over deploying unverified rules, which may either miss attacks or trigger on benign traffic.
The synthesized detection specification is used only during CEGIS-based verification and is not embedded in the final rule. Instead, verification establishes that the identified vulnerable parameter is consistent with the observed attack behavior.

\subsection{Rule Compilation}

Once a specification is verified, the compiler constructs a Suricata rule from the analysis results. Specifically, the compiler receives the original HTTP request and the LLM's analysis result, which includes the identified vulnerable parameter and attack characteristics, such as the payload syntax. It deterministically extracts the HTTP method, request path, attack value, and parameter location from the request, and maps the parameter to the appropriate Suricata matching buffer according to the request structure. For example, parameters are matched in the request body for POST requests and in the URI for GET requests.

To avoid overly broad signatures, generated rules are subjected to a rule-specificity test. Rules that match only generic request properties, such as the HTTP method and path, are discarded because they are likely to trigger on benign traffic. The remaining rules are emitted as verified Suricata rules. Since rule compilation is entirely deterministic, identical inputs always produce identical outputs without additional LLM-based repair or syntax correction.
Implementation details, prompts, source code, and additional rule generation examples are publicly available.\footnote{\url{https://github.com/mm-doshisha/CIKM2026-iot-vuln-pipeline}}
\begin{table}[t]
\centering
\caption{Attack Dataset Used for Suricata Rule Generation}
\label{tab:dataset}
\begin{tabular}{lrr}
\toprule
Vulnerability class & Count & Ratio (\%) \\
\midrule
Command injection & 132 & 47.0 \\
Other & 51 & 18.1 \\
Authentication bypass & 45 & 16.0 \\
Information disclosure & 30 & 10.7 \\
Path traversal & 23 & 8.2 \\
\midrule
Total attacks & 281 & 100.0 \\
\bottomrule
\end{tabular}
\end{table}

\section{Experiments}
\label{sec:experiments}

\subsection{Experimental Settings}
 
We collected 281 distinct CVEs, comprising 36 CVEs represented by attack traces from an open research dataset~\cite{torchlight} and 245 CVEs from the CISA Known Exploited Vulnerabilities catalog\footnote{\url{https://www.cisa.gov/known-exploited-vulnerabilities-catalog}}. For the 245 KEV vulnerabilities, we manually reconstructed each attack request from public exploit code or advisory descriptions. We selected only vulnerabilities affecting IoT or network devices that can be exploited through a single HTTP request. Table~\ref{tab:dataset} summarizes the resulting attack dataset, covering 49 vendors and five vulnerability categories.

To evaluate false positives under realistic deployment conditions, we used 281 benign communications in the UNSW-IoTraffic dataset~\cite{wannigama2025unsw}, collected from real IoT devices, using stratified sampling over device types and HTTP methods.
For specification synthesis, representative benign parameter values were calculated using CICIoT2023~\cite{ciciot23} in addition to UNSW-IoTraffic. CICIoT2023 provides benign body-parameter values that are insufficiently represented in UNSW-IoTraffic. To avoid data leakage, the 281 evaluation samples were excluded from this calculation. Although the synthesis and evaluation sets originate from the same datasets and may share some parameter values, the proposed method restricts rule matching to specific endpoints and parameter names, reducing the risk that such overlap directly affects evaluation on benign traffic.

We compare the proposed method against GRIDAI~\cite{gridai}, RuleXploit~\cite{rulexploit}, Moreno~\cite{moreno}, CMIRGen~\cite{cmirgen}, and AutoCombo~\cite{autocombo}. Methods other than GRIDAI and the proposed method used the auxiliary information obtained for each attack. All LLM-based methods, including the proposed method, used Qwen3-8B-BF16 running locally. Each method was evaluated on 281 attacks over three random seeds, yielding 843 attack instances per method. Results were averaged over the three random seeds. Generated rules were evaluated using Suricata 6.0.4.

\subsection{Detection and False-Positive Performance}

We evaluate generated rules using detection rate (DR) and false positive rate (FPR), both measured at the actual Suricata triggering layer. DR is the percentage of attack instances detected by generated rules. Attacks for which no rule can be generated are counted as misses. FPR is the percentage of benign IoT communications that incorrectly trigger generated rules and directly reflects operational false alarms.
DR is evaluated using the same attack traces from which rules are generated. Although this corresponds to evaluation on the generation data rather than unseen attacks, the objective of IDS rule generation is to construct rules that correctly detect the given attack when deployed in Suricata. Therefore, DR measures the success rate of rule generation rather than generalization performance. Similar evaluation protocols are commonly adopted in prior work~\cite{rulexploit,syrius,moreno}. 

\begin{table}[t]
\centering
\caption{Detection Rate (DR) and False Positive Rate (FPR) for All Methods}
\label{tab:main_results}
\begin{tabular}{llrr}
\toprule
Method & Category & DR (\%) & FPR (\%) \\
\midrule
Proposed & LLM Specification + Compiler & \textbf{81.5} & \textbf{0.0} \\
RuleXploit~\cite{rulexploit} & Direct LLM rule generation & 72.7 & 5.3 \\
GRIDAI~\cite{gridai} & Direct LLM rule generation & 69.5 & 0.5 \\
CMIRGen~\cite{cmirgen} & Signature generation & 54.9 & 10.3 \\
AutoCombo~\cite{autocombo} & Signature generation & 40.9 & 4.3 \\
Moreno+~\cite{moreno} & Direct LLM rule generation & 14.7 & 0.0 \\
\bottomrule
\end{tabular}
\end{table}

Table~\ref{tab:main_results} shows the results of all methods.
The proposed method achieves a DR of 81.5\% while maintaining an FPR of 0.0\%, making it the only method that simultaneously achieves high detection performance and zero false positives among all evaluated approaches. The remaining 18.5\% of attack instances correspond to cases where no Suricata rule could be generated.
RuleXploit and GRIDAI achieve competitive detection rates, but incur substantially higher false-positive rates because generated rules often match broader request patterns. In contrast, Moreno achieves an FPR of 0.0\% but detects only 14.7\% of attacks. CMIRGen and AutoCombo rely on benign data during rule generation; when evaluated on previously unseen benign traffic, both methods exhibit substantially lower detection rates and higher false-positive rates than the proposed method.

The low FPR of the proposed framework can be attributed to deterministic rule construction. Generated rules restrict matching to specific endpoints and vulnerable parameters and apply matching conditions derived from the observed attack value only within those contexts. Consequently, unrelated benign traffic rarely satisfies the matching conditions.

\begin{table}[t]
\centering
\caption{Ablation Study Results}
\label{tab:ablation}
\begin{tabular}{lrr}
\toprule
Configuration & DR (\%) & $\Delta$DR (percentage points) \\
\midrule
Full system (proposed) & \textbf{81.5} & --- \\
w/o CEGIS loop & 41.3 & -40.2 \\
w/o counterexample diagnosis & 57.8 & -23.7 \\
w/o compiler & 70.8 & -10.7 \\
\bottomrule
\end{tabular}
\end{table}

\subsection{Ablation Study}

Table~\ref{tab:ablation} reports the contribution of individual components. Removing the CEGIS refinement loop causes the largest performance degradation, reducing DR from 81.5\% to 41.3\%. This result indicates that attack understanding alone is insufficient for generating effective rules; iterative refinement using counterexamples is the main contribution of the proposed method.
Removing counterexample diagnosis reduces DR to 57.8\%, suggesting that counterexamples are useful not only for rejecting incorrect specifications but also for guiding synthesis toward more discriminative attack descriptions. Notably, counterexample diagnosis accounts for a substantial portion of the performance gain provided by the CEGIS-based refinement process.
The compiler also contributes substantially, reducing DR by 10.7 percentage points when removed. In this configuration, the LLM generates Suricata rules directly from the analysis result, bypassing rule compilation. Interestingly, none of the ablations increase deployment-time FPR, which remains 0.0\% in all cases. This observation suggests that the synthesis-verification loop primarily contributes to improving detection performance, while rule compilation provides deterministic and operationally consistent rule construction.
Overall, the ablation results indicate complementary roles within the framework: counterexample-guided verification improves attack detection, while rule compilation supports consistent rule generation.

\subsection{Failure Analysis}

We analyze why DR does not reach 100\% even when evaluated on the same attacks used for rule generation.
The failure modes differ across methods. The LLM-based baselines (RuleXploit, GRIDAI, and Moreno) generate rules for most attacks, but many of these rules fail to trigger in Suricata because of incorrect matching strings or buffer specifications. In contrast, the primary failure mode of CMIRGen and AutoCombo is the inability to generate rules for attacks that lack discriminative tokens or attributes. The proposed method exhibits a different behavior: almost all generated rules function correctly, and failures occur primarily when no valid detection specification can be synthesized. Specifically, there are attacks for which CEGIS fails to identify a specification that matches the attack value while rejecting all benign values.

Failure rates vary across vulnerability classes. 
Command-injection and path-traversal attacks are usually detected because they contain distinctive syntactic patterns such as shell metacharacters or traversal strings. In contrast, authentication-bypass and information-disclosure vulnerabilities often lack distinguishing request-level features. Since the proposed framework relies solely on HTTP requests, attacks whose maliciousness depends primarily on server-side state or response content remain difficult to characterize. Solving this limitation will be an important direction for future work.
\section{Conclusion and Future Work}

This paper presented a CEGIS-based framework for generating intrusion detection rules from HTTP request traces. Experiments on 281 real-world CVEs and 281 benign IoT communications demonstrated that the proposed method achieves an 81.5\% detection rate while maintaining an FPR of 0.0\%. The ablation study further showed that counterexample-guided verification is the primary contributor to detection performance, whereas rule compilation supports deterministic and operationally consistent rule construction.

Several directions remain for future work. First, while the current evaluation measures rule-generation success on observed attacks, future benchmarks should include multiple attack variants for the same vulnerability to evaluate generalization to unseen payloads and encodings. Second, extending verification with real devices or emulation environments~\cite{firmae} may enable detection of attack classes whose maliciousness depends primarily on server-side state, such as authentication bypass and information disclosure vulnerabilities. Finally, the current framework assumes attack traffic collected from honeypots or monitoring points; evaluating its robustness in environments where attack and benign traffic are mixed is an important direction for practical deployment.

\section{GenAI Usage Disclosure}

The authors used ChatGPT and Claude to assist with manuscript editing, language translation, and code development. All outputs produced by these tools were verified and modified by the authors. ChatGPT was also used to generate decorative icon images that appear in Figure~\ref{fig:overview}. The generated icons were used only for visualization purposes.
The authors are solely responsible for the manuscript content, methodology, experimental results, and conclusions presented in this paper.

\bibliographystyle{ACM-Reference-Format}
\bibliography{refs}

\appendix

\end{document}